# NUMERICAL RELATIVITY AND BLACK HOLE HORIZONS[*]

JOAN MASSÓ[1,2]

[1] *National Center for Supercomputing Applications (NCSA)*
*605 E. Springfield Av., Champaign, IL 61820, USA*

[2] *Departament de Física, Universitat de les Illes Balears*
*E-07071 Palma de Mallorca, SPAIN*

ABSTRACT

I report on recent progress in the exciting field of Numerical Relativity, with special attention to black hole horizons.

## 1. Introduction

After almost 80 years since its publication, the Einstein equations remain being one of the most beautiful and elusive systems of equations ever "created"/"discovered". The simple tensorial form that we all cherish leads to a set of 10 nonlinear, coupled, hyperbolic-elliptic partial differential equations that are not amenable to analytic study except in highly idealized cases involving symmetries or approximations. Their extreme complexity also poses a major challenge to numerical treatment. Until quite recently, Numerical Relativity (NR) has been restricted to 1D and 2D studies. In spite of many years of study, the solution space of the complete set of the Einstein equations is very much unknown. Fortunately, NR is finally coming of age. On the one hand, the development of massively parallel computers allows one to treat 3D problems, which were completely intractable several years ago due to limited computer power. On the other hand, a critical mass has been reached, with a growing community of research groups that in recent years has made significant progress[1].

A new generation of high performance, massively parallel computers, such as the NCSA Connection Machine CM-5 and others like it, promises to revolutionize computational physics, and Numerical Relativity, during this decade. We have now speeds of tens of Gigaflops and many Gigabytes of memory at our disposal, which is 100 times the power attainable just a few years ago on a Cray Y-MP, with much more to come in the next few years. To take advantage of such machines for NR, physicists and computer scientists are beginning to work together to develop new algorithms to solve the Einstein equations on parallel computers. Recently, the American National Science Fundation has funded a Grand Challenge Alliance at eight institutions with the goal of developing numerical codes to solve the problem of the 3D spiraling

---

coalescence of two black holes. This is a fundamental problem in relativity and astrophysics, as colliding black holes are among the most promising sources for generating gravitational waves that *will* be detected around the turn of the century[2].

In this paper, I give a partial overview of the work being performed at one of these institutions; NCSA. I also show some recent results in Numerical Relativity and, specifically, in the study of black hole horizons.

## 2. Numerical Relativity at NCSA

**2D Black Holes:** In the last years, the NCSA group has studied extensively axisymmetric black holes. First, the problem of a black hole interacting with a strong gravitational wave (Brill wave), where the wave collapses into the hole, exciting the so-called quasinormal oscillation modes of the hole[3,4]. This work has been extended to the cases of rotating black holes[5] and the head-on collision of two black holes[6]. The calculations show that for the head-on collision of two equal mass black holes, the quasinormal modes of the final black hole resulting from the collision are excited, and that the total energy released in the form of gravitational waves is on the order of 0.1% of the total mass of the system.

**3D codes:** Recently, we have developed two independent 3D codes in cartesian coordinates to study the most general possible spacetimes. One of these codes uses the standard $3+1$ ADM formalism with arbitrary shift and slicing conditions, and the other uses a 3D harmonic formulation[7] that puts the equations in an explicitly hyperbolic, first order, flux-conservative form. Both codes are very well suited to parallel machines. The harmonic code is one of the best performers of any application running on the NCSA 512 node CM-5, presently achieving nearly 20 GFlops. It has also been ported to most of the current supercomputer architectures (SGI Power Challenge, Convex Exemplar, IBM SP-2, Cray T3D, etc), with excellent results. These codes are now being used for 3D studies of black holes and gravitational waves.

**3D Black holes:** In order to make progress towards the Grand Challenge goal, the study of 3D black holes is fundamental. Unfortunately, these calculations are numerically difficult because of the sharp peaks of the metric functions that develop when evolving black holes with the use of a singularity avoiding time slicing[8]. We have tested many different slicing conditions for a single 3D black hole evolution. Geodesic slicing is used as a code test, since the solution is known analytically to crash at $t = \pi M$ ($M$ is the mass of the black hole) and it is always fun to make a code crash. With current resolutions (a grid size of about $200^3$), limited by computer memory sizes, we show that with certain lapse conditions we can evolve the black hole to about $t = 50M$ with errors within about 5% throughout the evolution[9]. We are now starting to study two black hole data sets.

**3D Waves:** There are theoretical, observational, and technical reasons to study pure wave spacetimes. This important area of research is for the most part uncharted territory. Previous analytic and numerical work on pure gravitational wave spacetimes, done in 1D and 2D, has led to many interesting discoveries, such as the formation of singularities from colliding plane waves or the existence of critical behavior in black hole formation in axisymmetric spacetimes. These discoveries raise interesting questions about what will happen in the generic 3D case. We have developed both fully nonlinear and linearized versions of both of our codes, so that we can compare differences and sort out nonlinear physics and code effects. We also use Fourier analysis to see how the codes propagate various frequency components. At present, our codes have passed all testbeds devised[10], and we are in the beginning phase of using the codes to investigate new physics. In Fig. 1 we show the evolution of an isosurface of the 3d metric function $g_{xx}$ for an initially linear 3D gravitational wave evolved with the harmonic code. The images are snapshots from a virtual reality environment (CAVE) application to analyze 3D spacetimes that we are developing[11] in collaboration with the Electronic Visualization Laboratory at Chicago.

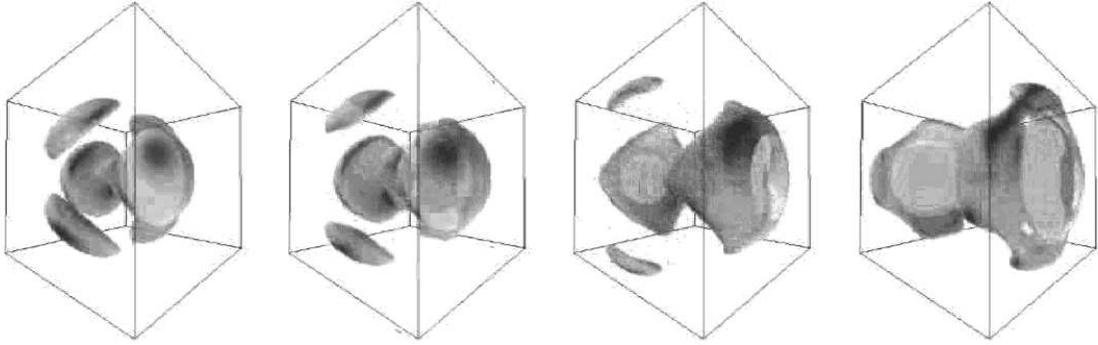

Figure 1: CAVE snapshots of the evolution of gravitational wave component $g_{xx}$.

**New Age:** In collaboration with the Relativity Group at the University of the Balearic Islands, we have developed a new formulation of the Einstein equations that casts them in an explicitly first order, flux-conservative, hyperbolic form. As an extension of previous work[7], we have shown how this now can be done for a wide class of time slicing conditions, including maximal slicing, making it potentially very useful for NR. This development permits the application to the Einstein equations of advanced numerical methods developed to solve the fluid dynamic equations, without overly restricting the time slicing, for the first time. We have obtained the full set of characteristic fields and speeds[12] and we are extending the harmonic 3D code to use this new formulation.

**Apparent Horizon Boundary Conditions:** We have studied how the numerical difficulties related with spacetime singularities in NR can be avoided by excising a region of the computational domain from inside the apparent horizon[8]. We have developed a scheme that is based on using (*i*) a horizon locking coordinate which locks the coordinate system to the geometry, and (*ii*) a finite differencing scheme which respects the causal structure of the spacetime. With this technique a spherically symmetric black hole can be evolved accurately well beyond $t = 1000M$, where $M$ is the black hole mass[13]. Now we are working on the 3D extensions.

**Other Projects:** We are working on 3D General Relativistic Hydrodynamics

and 3D boson stars. We have developed an efficient 3D apparent horizon finder[14] and we are investigating the use of Adaptative Mesh Refinement[15] in NR. We are also actively collaborating with computer scientists at the University of Illinois involved in the analysis and optimization to get the maximum performance of our parallel codes and of our linear systems solvers, and the use of constraint enforcers, finite elements, spectral methods, etc. Finally, we actively pursue the development of visualization techniques, including virtual reality, new interfaces using the World Wide Web, production of movies, etc.

## 3. Black Hole Horizons

Finally, I want to focus on one of the most exciting projects that we are currently working on: the study of black hole horizons. Indeed, the essential characteristics of a black hole in relativity are its horizons, in particular, the apparent horizon (AH) and the event horizon (EH). The membrane paradigm[16] characterizes black holes by the properties of its EH, which is regarded as a 2D membrane living in a 3D space, evolving in time and endowed with many everyday physical properties like viscosity, conductivity, entropy, etc. This provides a powerful tool in obtaining insight into the numerical studies of black holes, once the horizons can be determined for numerically constructed spacetimes. Most studies in NR on black hole horizons are in terms of the AH as it is a local object, readily obtained in a numerical evolution by searching a given time slice for closed surfaces whose outgoing null rays have zero expansion. In contrast, the EH is a globally defined object (the boundary of the causal past of the future null infinity) which makes it harder to find in numerical studies. Its location, and even its existence cannot be determined without knowledge of the complete four geometry of the spacetime. We have recently developed[17] a powerful and efficient method to locate the EH in NR for many dynamical spacetimes of interest, opening up the possibility of studying the dynamics of event horizons in NR.

There are two basic ideas in our methods in finding the horizon: (1) Integrating backward in time. We note that as going forward in time, outward going light rays emitted just outside the EH will diverge away from it, escaping to infinity, and those emitted just inside the EH will fall away from it, towards the singularity. This implies that in the time-reversed problem, any outward going photon that begins near the EH will rapidly be *attracted* to the horizon if integrated *backward* in time. In integrating backwards in time, the initial guess of the photon does not need to be very precise as it will converge to the correct trajectory after only a short time. We built our first generation of EH finder based on this backward photon method. Our second generation of EH finder has another important ingredient: (2) Rather than independently tracking all individual photons starting on a surface, we follow the entire null surface itself, as the EH is generically a null surface. A null generator of the null surface is guaranteed to satisfy the geodesic equation. A null surface defined by $f(t, x^i) = 0$ satisfies the condition $g^{\mu\nu} \partial_\mu f \partial_\nu f = 0$. Hence the evolution of the surface

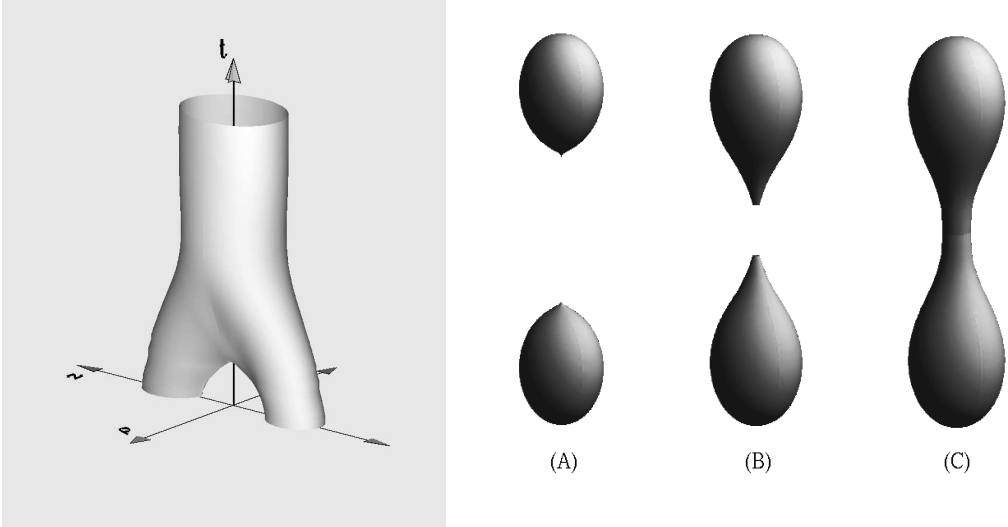

Figure 2: In the left figure, the geometric embedding of the event horizon for two black holes colliding head on is shown (it corresponds to an initial proper separation of $8.92M$, where $M$ is the mass of each hole). The $z$ coordinate marks the symmetry axis, and $t$ is the coordinate time. Initially the two holes are separate, and they coalesce into a single black hole. In the right figures, the time sequence (A), (B), (C) shows the merging with normalized pictures of the spatial surfaces.

can be obtained by a simple integration,

$$\partial_t f = \frac{-g^{ti}\partial_i f + \sqrt{(g^{ti}\partial_i f)^2 - g^{tt}g^{ij}\partial_i f \partial_j f}}{g^{tt}} \quad (1)$$

This method is simpler, more accurate and performs better in handling caustic structures of the EH than the backward photon method. Using this method we are able to trace accurately the entire history of the EH in a very short period of time. It takes just a few minutes to trace the EH on a computer workstation for an axisymmetric black hole spacetime. We contrast both our backward photon and backward surface methods with another method[18] that uses inefficient forward integration of individual photons to find the EH.

In Fig. (2) we show a geometric embedding of the coalescing horizons for the head-on collision of two black holes[6] with a normalized detail of how they merge. The embedding, which preserves the proper surface areas of the horizons, shows not just the topology but also the geometric properties of the horizon. Although such a picture of the embedding is familiar, this is the first time it has actually been computed. Currently we are actively working in the analysis of the horizon physics[19], with the hope of answering some of the open questions in horizon dynamics and unveiling the mystery of black holes.

## 4. Conclusion

As this paper is being written in November of 1994, by the time it is published it will be totally outdated. Fortunately, the magic of the world wide web will allow me to keep things updated. Contact http://jean-luc.ncsa.uiuc.edu/ for a repository of all our papers, movies, projects and exhibits, as well as links to other groups working in NR.

The future of Numerical Relativity is brilliant. General 3D gravitational systems, with their full nonlinearities, will finally be studied numerically within the next few years and the understanding of black hole horizons will crystalize. This opens the possibility of many interesting discoveries in physics of systems ranging from black holes to the universe as a whole.

## 5. Acknowledgements

The work that I have presented here involves many people and is a result of a collaboration between the NCSA group and other groups, notably the Washington University group. Thanks to all of them. I acknowledge the support of NSF grant PHY/ASC93-18152 (arpa supplemented) and a Fellowship (P.F.P.I.) from Ministerio de Educación y Ciencia of Spain.